\documentstyle[preprint,aps,psfig,epsf]{revtex}

\newcommand{\ave}[1]{\mbox{$\langle #1 \rangle$}}

\newcommand{\beq}{\begin{equation}}
\newcommand{\eeq}{\end{equation}}
\newcommand{\beqa}{\begin{eqnarray}}
\newcommand{\eeqa}{\end{eqnarray}}
\newcommand{\bmath}{\begin{mathletters}}
\newcommand{\emath}{\end{mathletters}}

\begin{document}
\draft
%\preprint{SISSA}
%%%COMMENT THE THREE LINES BELOW FOR 1 COLUMN and LINE JUST AFTER ABSTRACT
%\advance\textheight by 0.2in
%\twocolumn[\hsize\textwidth\columnwidth\hsize\csname@twocolumnfalse%
%\endcsname

\title{Macroscopic Quantum Fluctuations in the Josephson Dynamics \\
of Two Weakly Linked Bose-Einstein Condensates}
 
\author{Augusto Smerzi$^1$ and Srikanth Raghavan$^2$}
\address{
1) Istituto Nazionale di Fisica della Materia and  
International School for Advanced Studies,\\
 via Beirut 2/4, I-34014, Trieste, Italy,\\
2) Rochester Theory Center for Optical Science and Engineering and \\
Department of Physics and Astronomy, University of Rochester, Rochester 
NY 14627}
\date{\today}
\maketitle
\begin{abstract}
We study the quantum corrections to the Gross-Pitaevskii equation 
for two weakly linked Bose-Einstein condensates. 
The goals are: 1) to investigate dynamical regimes 
at the borderline between the classical 
and quantum behaviour of the bosonic field; 2) to search for new macroscopic 
quantum coherence phenomena not observable with other 
superfluid/superconducting systems. Quantum fluctuations renormalize 
the classical Josephson oscillation frequencies.
Large amplitude phase 
oscillations are modulated, 
exhibiting collapses
and revivals. We describe a new inter-well oscillation mode,  
with a vanishing (ensemble averaged) mean value of the observables,
but with oscillating mean square fluctuations. Increasing    
the number of condensate atoms, we recover 
the classical Gross-Pitaevskii 
(Josephson) dynamics, without invoking the symmetry-breaking of the Gauge
invariance.
\end{abstract}
\pacs{PACS: 74.50.+r,03.75.Fi}
%%%%%%COMMENT LINE BELOW FOR 1 COLUMN %%%%%%%%%%%%%%%%%%%%%%%%
%]

The experimental observation of the Bose-Einstein condensation
of a trapped, dilute gas of alkali atoms \cite{anderson95:misc}, 
and the high accuracy of the engineering
\cite{andrews97:misc,anderson98}, 
are opening a new avenue to investigate the interplay
between macroscopics and quantum coherence.
Foundational problems of 
quantum theory \cite{leggett91:misc,cirac98}
and condensed matter \cite{leggett91:misc,leggett95} 
can be addressed through real (and not just ``gedanken'') experiments;
dynamical regimes not accessible with other
superconducting/superfluid systems might be
testable.

The main goal of this work is to study the quantum corrections to the 
classical Gross-Pitaevskii dynamics \cite{pitaevskii61} of 
two weakly linked Bose-Einstein condensates (BEC's) forming 
a Josephson junction. 

The Gross-Pitaevskii equation (GPE) has been shown to describe 
quite accurately the dynamical regimes experimentally investigated  
so far \cite{dalfovo:condmat98}. On the other hand,
BEC's can be experimentally created with a number of atoms ranging
from few thousand to several millions, and in a wide variety of confining
geometries \cite{anderson95:misc,andrews97:misc,anderson98}. This 
is opening the important possibility of
 studying dynamical regimes at the borderline between the
quantum and classical nature of
 the bosonic field, and, more generally,
to search for new macroscopic quantum phenomena
not obervables with 
other superfluid/superconducting systems. 
 
Different Josephson dynamical regimes  
are characterized by the ratio of the ``Josephson coupling energy" $E_J$
and the ``on-site energy'' $E_s$ 
\cite{barone82,likharev91,tinkham96:misc,packard98:misc}.
In the limit 
$E_J \gg E_s$ (often referred in the literature
as ``classical'' \cite{likharev91,tinkham96:misc}), 
both the phase difference and 
the relative number of condensate atoms are well defined. Nevertheless, 
we will see that 
quantum corrections can significantly modify the classical dynamics even for 
$E_J/ E_s \sim 10^2 $, a regime accessible with current BEC technology.
Quantum fluctuations renormalize the classical Josephson
oscillation frequencies.
Large amplitude phase oscillations are modulated by partial collapses
and revivals.
It is well known
that 
the relative phase of two weakly linked
systems diffuses \cite{leggett91:misc,sols94,anderson84,imamoglu97}
and subsequently revives \cite{sols94} 
after the suppression of the Josephson coupling. 
In effect, even in two $coupled$ condensates, 
the relative phase
diffuses in the self-trapped, running-phase \cite{sfgs-rsfs,milburn97}
regime. 
Then, partial or complete
revivals could occur, due to the finite 
number of condensate atoms. An asymmetric potential can also induce
phase diffusion \cite{villain99}.  

In the limit of large condensates, 
the dynamical equations for the mean values of the 
physical observables decouple from the equations governing
the respective quantum fluctuations, with a   
smooth crossover from the quantum 
to the classical GPE regime.
 
The classical boson Josephson junction (BJJ) equations, derived by the
GPE in the ``two-mode'' approximation
\cite{sfgs-rsfs,milburn97,villain99,zapata98,williams99},
can be cast in terms of two canonically conjugate variables:
the relative population $N$ and phase $\phi$ between the two traps. 
Quantizing BJJ, the 
c-numbers $N$ and $\phi$ are replaced by the corresponding 
operators, 
satisfying the commutation relation 
$[\hat{\phi},\hat{N}] = i$  \cite{packard98:misc,nota:phasequant}. 
Then the Hamiltonian of two 
weakly coupled condensates reads
\cite{nota:nonlinearcorrect}:  
\beq
\hat{H} = {E_s \over 2} \hat {N}^2 - E_J  \cos \hat{\phi} + 
\Delta E_0 ~\hat{N}
\label{eq:eq1}
\eeq
where $E_J~(\sim N_T^\alpha;~\alpha \sim 1)$ is the 
``Josephson coupling energy''; 
$E_s~(\sim N_T^{- \beta}$, with $\beta = 3/5$ in $3d$ traps)
is the ``on-site energy'', the analog of the charging energy
in a superconducting Josephson Junction (SJJ); 
$N_T$ is the total number of condensate atoms.
$\Delta E_0$ is the zero-point energy difference in two
asymmetric traps \cite{sfgs-rsfs}, or an applied chemical potential
difference 
(induced, for example, by the gravitational potential  
in vertical traps \cite{anderson98}).
The coefficients $E_J, E_s$ are determined by the BJJ geometry
and the total number of atoms. They can be (consistently) calculated
as overlap integrals of two orthogonal 
one-body Gross-Pitaevskii wave functions \cite{sfgs-rsfs}.

In the phase representation, the operators in Eq.~(\ref{eq:eq1})
are expressed as
$\hat{N} = - i \frac{\partial}{\partial \phi}, 
\hat \phi = \phi$.  
Then the dynamical equation for the amplitude $\Psi(\phi,t)$ 
is $(\hbar = 1)$: 
\beq
i \frac{\partial \Psi(\phi,t)}{\partial t} = 
-\frac{\partial^2 \Psi (\phi,t)}{\partial \phi^2}  -
\Gamma \cos \phi ~\Psi(\phi,t) 
-i E_0 \frac{\partial \Psi(\phi,t)}{\partial \phi} ,
\label{eq:eq2}
\eeq
where $\Gamma = {2 E_J \over E_s}$, $E_0 = {{2 \Delta E_0} \over E_s}$ 
and the time has been rescaled as
${E_s \over 2} t \to t$. 
Since we are considering an isolated, energy conserving system, the 
``potential''
is periodic and defined on a $2 \pi$-ring, with the wave-function boundary 
conditions $\Psi(\phi) = \Psi(\phi+2 \pi)$. 

In the context of SJJ's, (where only {\em stationary} 
regimes are experimentally
accessible),  
Eq.~(\ref{eq:eq2}) is the {\em drosophila}
 of low-capacitance systems,
\cite{barone82,likharev91,tinkham96:misc}. 
The effect of dissipation on their quantum statistical 
properties (like phase transitions from normal to
superconducting phases) has been extensively studied \cite{schon90}.
Other typical effects include the renormalization of the critical Josephson
current \cite{panyukov87}, and the macroscopic quantum tunneling 
among metastable minima of the "washboard" potential 
\cite{barone82,likharev91,tinkham96:misc}.

In an BJJ,
on the other hand, {\em dynamical} density oscillations can be studied 
by shifting the position of the laser barrier 
or tailoring the traps \cite{andrews97:misc,anderson98}. 
(A similar argument holds when considering Raman 
transitions between two condensate
 in different hyperfine levels of a single
traps). The small frequency, $ < $ kHz, oscillations of the 
population imbalance, and
the corresponding mean square deviations, 
might be directly monitored by
 destructive or non-destructive techniques. 
It is worth stressing, then, that the set of experimentally
accessible observables in BJJ is rather different from the SJJ one.
New collective quantum phenomena, not accessible with other
superconducting/superfluid systems, might be observed
in BEC's.  
 
In this Letter we study analytically 
the quantum corrections to
the classical Gross-Pitaevskii-Josephson equations,
providing a simple   
framework to study quantitatively 
a mesoscopic BJJ \cite{nota:mathieu}. 

We consider a time-dependent variational approach.
The time evolution of the variational 
parameters is characterized by the minimization of
an action with the effective Lagrangian:
\beq
L( q_i, \dot{q_i}) = i \ave{\Psi \dot{\Psi}} - \ave{\Psi \hat{H} \Psi}
\label{eq:eq4}
\eeq
with $\hat{H} =  -\frac{\partial^2 }{\partial \phi^2}  - \Gamma \cos \phi 
- i E_0 \frac{\partial }{\partial \phi}$
and $\Psi(\phi, q_i(t))$, $q_i$ being the time-dependent parameters.
This provides the familiar Lagrange equations:
\beq
\frac {d } {dt} \frac{\partial L }{\partial \dot{q_i}} =
 \frac {\partial L}{\partial q_i}
\label{eq:eq5}
\eeq
We choose the time-dependent variational wave function as:
\beq
\Psi(\phi,t) = R[\phi-\phi_c(t),\lambda(t)] 
~e^{i S[\phi-\phi_c(t), p(t), \delta(t)]}
\label{eq:eq6}
\eeq
with the square root of the probability density 
$R$ and the dynamical phase $S$ being real, and:  
\beq
S = p(t)~ [\phi - \phi_c(t)] + {\delta(t) \over 2}~ [\phi - \phi_c(t)]^2.
\label{eq:eq7}
\eeq
The pairs of time-dependent parameters 
$\phi_c(t),p(t)$ and 
$\lambda(t),\delta(t)$ are 
canonically conjugate:
\bmath
\beqa
\dot{\phi_c} &=& {{\partial H}\over {\partial p}} = 2 p + E_0\\
\dot{p} &=& -{{\partial H}\over {\partial \phi_c}} = 
-{{\partial }\over {\partial \phi_c}} \ave{V(\phi)}\\
\dot{\lambda} &=& {{\partial H}\over {\partial \delta}} = 
4 \lambda \delta\\
\dot{\delta} &=& - {{\partial H}\over {\partial \lambda}} = - 2 \delta^2
+ {{\partial }\over {\partial \lambda}} 
[\int_{\phi_c - \pi}^{\phi_c +\pi}{R R'' d\phi} - \ave{V(\phi)}]
\label{eq:eq33}
\eeqa
\emath
with the effective Hamiltonian:
\beq
H = \ave{T} + \ave{V} = 
p^2 + 2 \lambda \delta^2 -  \int_{\phi_c - \pi}^{\phi_c + \pi}
{R R'' d\phi} + \ave{V(\phi)} + E_0~ p.
\label{eq:eq34}
\eeq
Thus $p(t)$ is the momentum associated 
with the center of mass motion 
$\phi_c(t) = 
\int_{\phi_c - \pi}^{\phi_c + \pi} R^2(\phi - \phi_c) 
~\phi~ d \phi$, and $\delta(t)$ is the conjugate
momentum of the width of the 
wave-function, which is proportional to 
$\lambda(t) = {1 \over 2} \int_{\phi_c - \pi}^{\phi_c + \pi} 
R^2 ~\phi^2~d \phi $.
The $\langle...\rangle$ means 
$\int_{\phi_c - \pi}^{\phi_c + \pi} |\Psi(\phi - \phi_c)|^2~...~d \phi$,
and the prime $'$ stands for ${\partial \over {\partial \phi}}$.
The mean value of the population imbalance between the two traps is
$N(t) = \ave{\Psi(\phi,t) \hat{N} \Psi(\phi,t)} = p(t)$, 
the relative phase is
$\phi(t) = \phi_c(t)$,
and the corresponding mean square deviations are 
$\sigma^2_N(t) = \ave{\hat{N^2}} - \ave{\hat{N}}^2 = \ave{T(t)}$
and $\sigma^2_\phi(t) = 2 \lambda(t)$.

For $\Gamma \gg 1$,  $R[\phi-\phi_c(t),\lambda(t)]$ can be well
approximated by a Gaussian: 
\beq
R(\phi,t) = (4 \pi \lambda)^{-1/4} 
e^{-{1 \over {8 \lambda}} (\phi - \phi_c)^2}
\label{eq:eq8}
\eeq
with the caveat that during the dynamics 
its width $2 \sigma_\phi(t) = 2 \sqrt{2 \lambda(t)} \ll 2 \pi$.

The equations of motion become:
\bmath
\label{eq:eq11}
\beqa
\dot{N} &=& -\Gamma \sin \phi~ e^{-{\sigma_{\phi}^2 \over 2}} \\
\dot{\phi} &=&  2 N + E_0\\
\dot{\sigma}_{\phi} &=& 2 {\sigma}_{\phi} \delta \\
\dot{\delta} &=& - 2 \delta^2
+{1 \over {2 \sigma_{\phi}^4}} - \Gamma \cos \phi~
e^{- {\sigma_{\phi}^2 \over 2}}
\eeqa
\emath
with the total (conserved) energy and the relative population dispersion:
\beqa
H &=& N^2 + \sigma_N^2 - \Gamma \cos\phi~ e^{- {\sigma_{\phi}}^2 \over 2}
+ E_0 ~N \\
\sigma_N &=& {1 \over {2 \sigma_{\phi}}}
\sqrt{1 + 4 \sigma_\phi^4 \delta^2}
\label{eq:eq11b}
\eeqa
The canonically conjugate dynamical variables are $N, \phi$, as in the 
classical Josephson Hamiltonian, 
and the pair ${\sigma_\phi^2 \over 2}, \delta= {1 \over \sigma_\phi}
\sqrt{\sigma_N^2 - {1 \over {4 \sigma_\phi^2}}}$, 
characterizing the respective quantum fluctuations. 
As expected, $\sigma_N \sigma_{\phi} \ge {1 \over 2}$ 
during the dynamics. The classical Josephson equations
are recovered
from Eqs.(\ref{eq:eq11}~a,b) in the 
limit $\sigma_{\phi} \to 0$. We will discuss more about the
transition from the quantum to the classical regimes in the following.
  
The variational ground state energy of Eq.~(\ref{eq:eq2}) is given by:
\beq
E_{gs} = {1 \over {4 \sigma_{\phi,s}^2}} - \Gamma
e^{- \sigma_{\phi,s}^2 /2} + {E_0^2 \over 2}  = 
\sigma_{N,s}^2 - \Gamma
e^{- 1 / 8 \sigma_{N,s}^2} - {E_0^2 \over 4}
\label{eq:eq12}
\eeq
where $\sigma_{\phi,s}, \sigma_{N,s}$ 
are the solution of: 
\beq
2 \sigma_{\phi,s}^4 ~e^{-\sigma_{\phi,s}^2 / 2} = 
e^{-\frac{1}{8\sigma_{N,s}^2}}/8  \sigma_{N,s}^2=
{1 \over \Gamma}
\label{eq:eq13}
\eeq
The stationary results were first discussed 
in \cite{tinkham96:misc},
where Eq.s~(\ref{eq:eq12},\ref{eq:eq13}) were obtained minimizing the 
ground state energy with a Gaussian trial wave function in the case $E_0 = 0$.
Linearizing Eq.~(\ref{eq:eq11})
for small amplitude $\phi$ oscillations, we have:
$\ddot{\phi} = - 2 \Gamma e^{- \sigma_{\phi,s}^2 / 2} \phi$ 
The condensate atoms oscillate coherently with a frequency 
(unscaled):
\beq
\omega_q = \sqrt{E_s E_J} ~e^{- \sigma_{\phi,s}^2 /4} 
\label{eq:eq15}
\eeq
where the classical Josephson relation gives $\omega_c = \sqrt{E_s E_J}$.
The quantum fluctuations renormalize the Josephson plasma frequency,
with ${\omega_q \over \omega_c} = \exp(-\sigma_{\phi,s}^2 / 4) = 
(\sigma_{\phi,s}^2 \sqrt{2 \Gamma})^{-1}$.
Notice that in the linear regime, the current-phase Eqs.(\ref{eq:eq11}~a,b)
are effectively 
decoupled from the dynamics of the respective fluctuations
Eqs.(\ref{eq:eq11}~c,d). On the contrary, 
for large amplitude $\phi$ oscillations, 
Eqs.~(\ref{eq:eq11}) cannot be decoupled.
 In this case the exponential factor 
modulates the amplitude and the frequencies of the oscillations,
inducing partial
collapses and revivals. This can be seen in Fig.(1 a-d), where we
show the population imbalance, the relative phase and the respective 
mean square deviations as a function of time.  
 
Above the critical point ($N = 0,~\phi = \pi/2$),
the phase $\phi(t)$ starts running, Fig.(2b), and the system is set into
macroscopic quantum self-trapping (MQST) mode \cite{sfgs-rsfs,milburn97}. 
The width of
the wave function grows and the amplitude of oscillations 
`collapses', Fig.(2a). 
In the deep MQST regime, when $N(t) \simeq N(t=0)$,
the phase diffuses
as  $\sigma_{\phi}^2(t) \simeq \sigma_{\phi,s}^2
+ \frac{E_s^2}{4 \sigma_{\phi,s}^2}t^2$, Fig(2d),
regardless of the initial 
value of
$N(t=0)$ \cite{nota:dispersion}.
 The relative population oscillations collapse  
with a life time $\tau \simeq 2 E_J^{-{1 \over 4}} E_s^{-{3 \over 4}}$,
while the $\sigma_N^2(t)$ tends to a constant value, Fig(2c). 
However, since the total number of condensate atoms is finite, 
the phase can eventually revive partially or 
completely. This can be seen rewriting the wave function in the 
$N$ representation:
$\Psi(\phi,t) = \sum_N a_N \Phi_N(\phi) e^{- {i \over \hbar} E_N t}$,
where $E_N$ are the eigenenergies of Eq.~(\ref{eq:eq2}). 
For instance, in the limit $E_s N_T \ll E_J$, the eigenvalue
spectrum is approximately linear $E_N \sim {E_J \over N_T} N$, 
and the revival time is $\tau_R \sim h N_T / E_J$. More generally, 
the occurrence
of a complete or partial revival, or the complete
destruction of it, depends
on the detailed eigenspectrum of the Hamiltonian. We note
that Eq.~(\ref{eq:eq11}) cannot describe
the revival after a complete collapse 
since the 
Gaussian ansatz Eq.~(\ref{eq:eq8}) (and, consequently, the semiclassical
approximation underlying it) breaks down when
$\sigma_{\phi} \simeq \pi$.  

Eq.s~(\ref{eq:eq11}) admit, as a dynamical 
solution, a quite peculiar oscillation mode, with zero relative phase and
population imbalance, 
but oscillating fluctuations, according to:
\bmath
\label{eq:eq1111}
\beqa
N(t) &=& 0 \\
\phi(t) &=&  0 \\
\sigma_N^2(t) &=&  {1 \over{4 \sigma_\phi^2}} +  {\dot{\sigma_\phi}^2 \over4}\\
\ddot{\sigma}_\phi &=&  {1 \over \sigma_\phi^3} -  
2 \Gamma \sigma_\phi ~e^{- {\sigma_{\phi}^2 \over 2}} 
%{{d^2}\over {d t^2}} \sigma_\phi^2 & ?=? & {1 \over \sigma_\phi^2} 
%- 2 \Gamma \sigma_\phi^2 ~e^{- {\sigma_{\phi}^2 \over 2}} \;;\;
%{{d^2}\over {d t^2}} \sigma_\phi^2  ?=?  
%4 \Gamma (2 - \sigma_{\phi}^2)  e^{- {\sigma_{\phi}^2 \over 2}}
\eeqa
\emath
with initial conditions $N(t=0) = 0, \phi(t=0) =0;~E_0 =0$ and arbitrary
$\sigma_\phi(t=0)$. 
This collective oscillation mode
can be experimentally observed
by lowering  
the height of the barrier of a BJJ ensemble 
in thermodynamic equilibrium.
This corresponds to changing $\Gamma$ in Eq.~(\ref{eq:eq1111}) from its 
initial value.
The temporal evolution of the ensemble averaged observables 
and the respective mean square deviations
can be calculated by 
tracing the dynamical $N(t), \phi(t)$ trajectories of each junction. 

$Classical~limit$.
Increasing the number of atoms,
$\sigma_\phi \to 0$ as $\Gamma^{-{1 \over 4}} 
\sim N_T^{-{1 \over 4}(\alpha + \beta)}$. 
Moreover, for a given initial value $N(t=0)$, 
the amplitude of the ``particle'' oscillations in the 
$\phi$-potential of Eq.~(\ref{eq:eq2}), decreases as 
$\phi_{max} \sim \Gamma^{-{1 \over 2}} 
\sim N_T^{-{1 \over 2}(\alpha + \beta)}$.
Then Eqs.~(\ref{eq:eq11}~a,b)  
decouple from Eqs.~(\ref{eq:eq11}~c,d), and the time evolution 
of the mean values of current and phase become independent
of the corresponding dispersions.
In the MQST regime, the collapse time (and, consequently, 
the time over which the semiclassical predictions are reliable), 
increases as
$\tau \sim N_T^{{1\over 4} (3 \beta - \alpha)}$. 
In this framework the 
classical limit emerges naturally, without invoking any symmetry breaking 
argument, as discussed in \cite{leggett95}.
 
$Numerical~estimates$. 
Following the analytical estimations of the Josephson coupling energy 
and the on-site energy given for two 
weakly coupled condensates in \cite{zapata98}, we have:
$\Gamma \simeq 1.7 N_T {a_0 \over a_s} {{\exp(-S)} \over {\tanh(S/2)}}$,
with $a_0,a_s$ the trap length and the scattering length, respectively, and
with $S \sim {1 \over \hbar} \sqrt{2 m \sigma_B^2 (V_0 - \mu)}$.
$\sigma_B$ is the width of the barrier, $V_0$ its height, and
$\mu$ the chemical potential.  
For typical traps and condensates,
 $a_0 \sim 10^4~{\rm \AA},~a_s \sim 50~{\rm \AA}$ and
$\sigma_B \sim 5~\mu {\rm m}$. With a height of the barrier such that
$ (V_0 - \mu) \sim 10 ~nK$, we have $\Gamma \sim 10-100$ for $N_T \sim 1000$.
Varying the width and/or the height 
of the barrier, and the total number of condensate atoms, the system might span
from the $\Gamma \ll 1$ to the $\Gamma \gg 1$ limits.
The temperature
should be small compared to the Josephson coupling energy 
\cite{zapata98} to avoid destroying the
quantum fluctuations.
Damping effects are also reduced by decreasing the
total number of atoms. Such regimes (vanishing small temperatures, 
and small population per situ $\sim 1000$ in an optical array) are under
current investigations \cite{anderson98}. 
Concluding, we remark that Eq.s~(\ref{eq:eq11}) can be straightforwardly
generalized to
describe interwell tunneling in an array of trapped condensates. Work 
in this direction as well as on  
the quantitative analysis of the effects of
temperature and damping on the quantum dynamics 
are in progress. 

This work was partly supported by NSF Grant PHY94-15583.

%\bibliography{%

\begin{thebibliography}{10}
\bibitem{nota:nonlinearcorrect}
{Eq.~(\ref{eq:eq1}) holds for $N / N_T \ll 1$ and $E_s \gg E_J/N_T^2$ 
where $N_T$
  is the total number of condensate atoms
  The inclusion of
  nonlinear corrections has been studied in 
  \cite{sfgs-rsfs,milburn97}.}
\bibitem{nota:mathieu}
{Clearly, Eq.~(\ref{eq:eq2}) can be studied numerically, in terms of the
  eigenstates of the Mathieu equation or by direct integration; but this would
  obscure the subtle relations between observables in the semiclassical regime.
  }
\end{thebibliography}
%/u/fac/srirag/tex/bibtexbase/bec,%
%/u/fac/srirag/tex/bibtexbase/thesbib,%
%/u/fac/srirag/tex/bibtexbase/helium%
%}

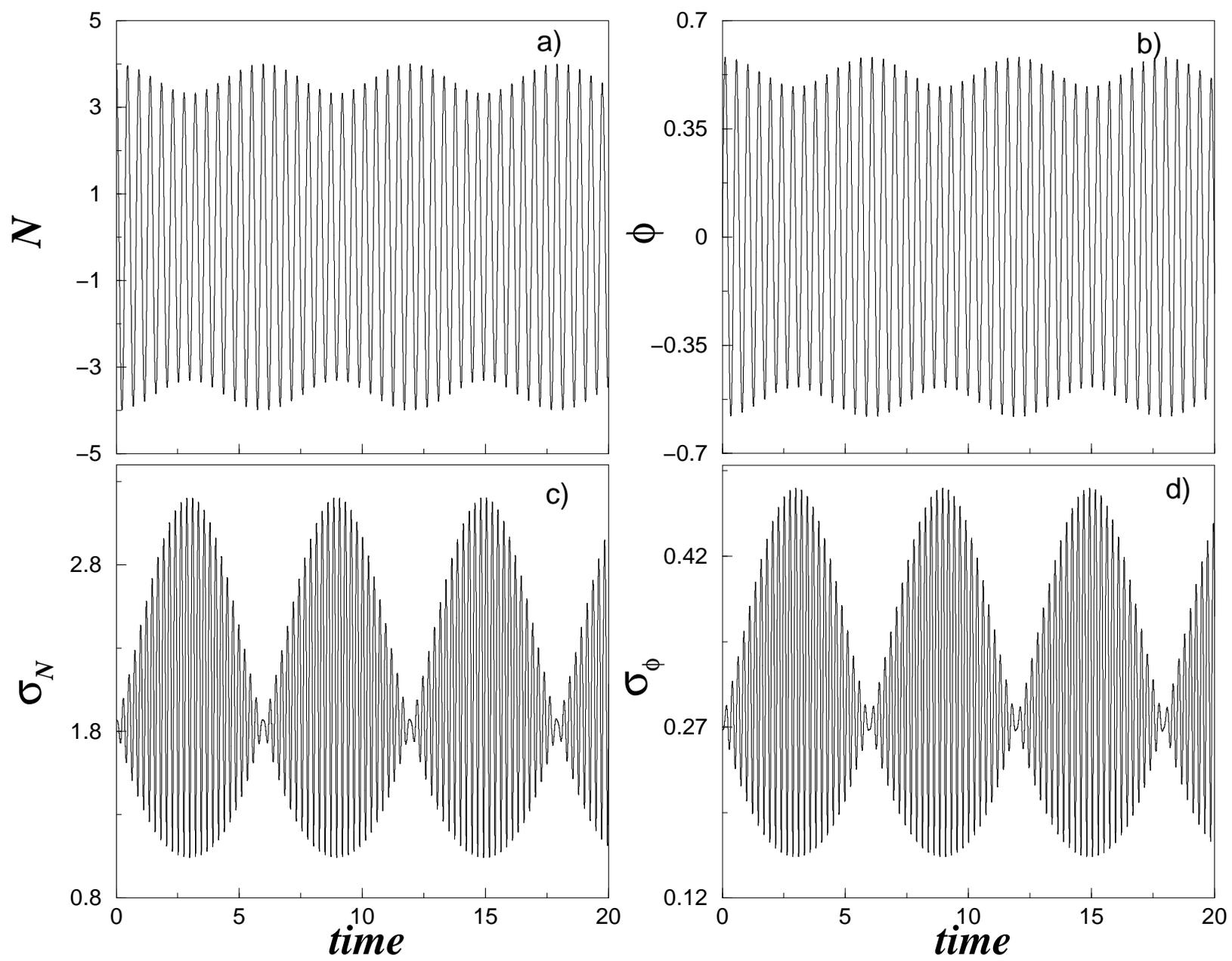
\begin{figure}
\caption{The population imbalance $N$, the relative phase $\phi$, 
and the corresponding
fluctuations $\sigma_N$ and $\sigma_\phi$ 
as a function of time. The initial conditions are
$N = 4,~\phi = 0$,~ $\Gamma =100,~\delta = 0,
~\sigma_{\phi} = 0.26$ and $E_0 = 0$}
\label{fig:fig1}
%%%FIGURE 1
\end{figure}
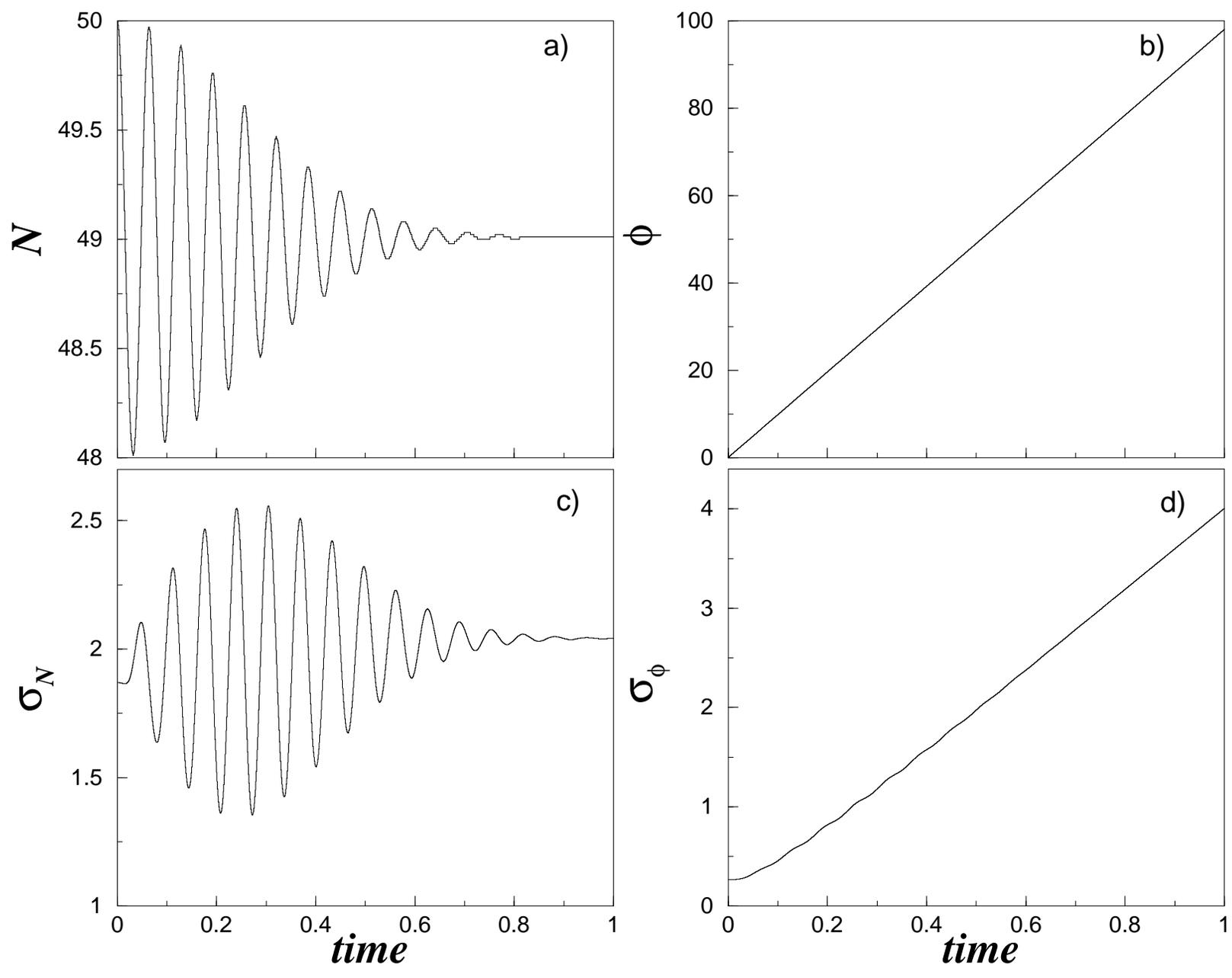
\begin{figure}
\caption{$N,\phi,\sigma_N,\sigma_\phi$
as a function of time. 
The initial conditions are the same as in Fig.(1) except
$N(t=0) = 50$. }%The oscillations
%are around a mean value $N(t) \ne 0$, corresponding to the macroscopic
%quantum self-trapping regime.}
\label{fig:fig2}
%%%FIGURE 1
\end{figure}

\end{document}